\documentclass{sf2a-conf}
\usepackage{graphicx}
\usepackage{natbib}
\begin{document}

\TitreGlobal{SF2A 2010}

\title{High-energy radiation from the relativistic jet of Cygnus X-3}
\author{Cerutti, B.}\address{Laboratoire d'Astrophysique de Grenoble, UMR 5571 CNRS, Universit\'e Joseph Fourier, BP 53, 38041 Grenoble, France}
\author{Dubus, G.$^1$}
\author{Henri, G.$^1$}
\runningtitle{High-energy radiation from the relativistic jet of Cygnus X-3}
\setcounter{page}{1}

\index{Cerutti, B.}
\index{Dubus, G.}
\index{Henri, G.}

\maketitle
\begin{abstract}
Cygnus X$-$3 is an accreting high-mass X-ray binary composed of a Wolf-Rayet star and an unknown compact object, possibly a black hole. The gamma-ray space telescope {\em Fermi} found definitive evidence that high-energy emission is produced in this system. We propose a scenario to explain the GeV gamma-ray emission in Cygnus X$-$3. In this model, energetic electron-positron pairs are accelerated at a specific location in the relativistic jet, possibly related to a recollimation shock, and upscatter the stellar photons to high energies. The comparison with {\em Fermi} observations shows that the jet should be inclined close to the line of sight and pairs should not be located within the system. Energetically speaking, a massive compact object is favored. We report also on our investigations of the gamma-ray absorption of GeV photons with the radiation emitted by a standard accretion disk in Cygnus X$-$3. This study shows that the gamma-ray source should not lie too close to the compact object.
\end{abstract}

\section{Introduction}

Cygnus X-3 is an accreting high-mass X-ray binary with relativistic jets, {\em i.e.} a microquasar. This system is composed of a Wolf-Rayet star (see {\em e.g.} \citealt{1996A&A...314..521V}) and an unknown compact object, probably a black-hole, in a tight 4.8~hours orbit
and is situated at about 7~kpc from Earth \citep{2009ApJ...695.1111L}. The gamma-ray space Telescopes {\em AGILE} \citep{2009Natur.462..620T} and {\em Fermi} \citep{2009Sci...326.1512F} have detected gamma-ray flares at GeV energies in the direction of Cygnus X$-$3 (a new gamma-ray flare was recently reported, see \citealt{2010ATel.2646....1C}). This identification is firmly established since the orbital period of the system was found in the {\em Fermi} dataset. This is the first
unambigous detection of a microquasar in high-energy gamma rays. 
The gamma-ray emission in Cygnus~X$-$3 is transient and correlated with powerful radio flares, associated with the presence of a relativistic jet and episodes of major ejections in the system. This feature suggests that the gamma-ray emission originates from the jet. In this proceeding, we present a model for the gamma-ray modulation in Cygnus X$-$3 (Sect.~\ref{sect_mod}, see \citealt{2010MNRAS.404L..55D} for more details). GeV photons could be absorbed by the soft X-rays emitted by an accretion disk around the compact object. We report also on our studies of the gamma-ray opacity in Cygnus~X$-$3 and put constraints on the location of the gamma-ray source in the system (Sect.~\ref{sect_abs}).

\section{Gamma-ray modulation}{\label{sect_mod}}

\subsection{The model}

The model relies on simple assumptions. Energetic electron-positron pairs are located at a specific altitude $H$ along the jet and symmetrically (with respect to the compact object) in the counter-jet. These acceleration sites could be related to recollimation shocks as observed in some AGN such as M87 \citep{2006MNRAS.370..981S}, possibly produced by the interaction of the dense Wolf-Rayet star wind and the jet. This possibility seems to be corroborated by recent MHD simulations (see \citealt{2010A&A...512L...4P}). Pairs are isotropic in the comoving frame and follow a power-law energy distribution.
The total power injected into pairs is $P_e$. The jet is relativistic (with a bulk velocity $\beta=v/c>0$) and inclined in an arbitrary direction parameterized by the spherical angles $\phi_j$ (polar angle) and $\theta_j$ (azimuth angle). The orbit of the compact object is circular with a radius $d=3\times 10^{11}~$cm. We define here the orbital phase $\phi$ such as $\phi\equiv 0.25$ at superior conjunction and $\phi\equiv 0.75$ at inferior conjunction. The Wolf-Rayet star (effective temperature $T_{\star}\sim 10^5~$K, stellar radius $R_{\star}\sim R_{\odot}$) provides a high density of seed photons for inverse Compton scattering on the relativistic pairs injected in the jet ($n_{\star}\approx 10^{14}~$ph~$\rm{cm}^{-3}$ at the compact object location). Because of the relative position of the star with respect to the energetic pairs and the observer, the inverse Compton flux is orbital modulated. This is a natural explanation for the orbital modulation of the gamma-ray flux observed in Cygnus X$-$3. Other sources of soft radiation ({\em e.g.} accretion disk, CMB) can be excluded to account for the modulation, and could instead contribute to the DC component. In addition to anisotropic effects, the relativistic motion of the flow should be considered in the calculation of the Compton emissivity (Doppler-boosting effects, see \citealt{2010A&A...516A..18D} for technical details).

\subsection{Results}

We applied this model to Cygnus~X$-$3 for two extreme orbital solutions as given in \citet{2008MNRAS.386..593S}. In the first solution (orbit inclination $i=30^{\rm{o}}$), the compact object is a 20~$M_{\odot}$ black hole orbiting a 50~$M_{\odot}$ Wolf-Rayet star. In the second solution ($i=70^{\rm{o}}$), the system is composed of a 1.4~$M_{\odot}$ neutron star and a 5~$M_{\odot}$ star.

\begin{figure}[h]
   \centering
   \includegraphics*[width=9cm]{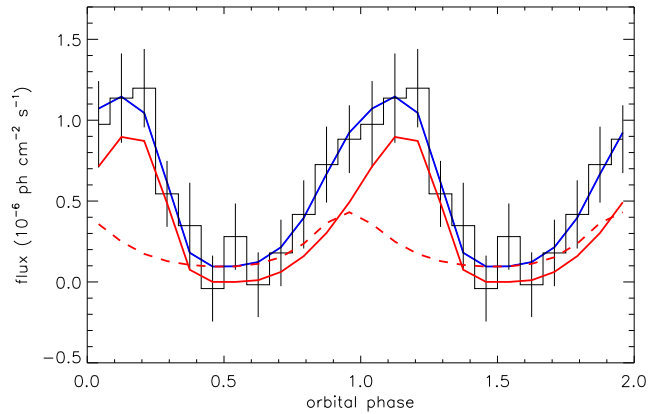}
   \caption{High-energy gamma-ray flux ($>100$ MeV) as a function of the orbital phase (two full orbits) for the black hole solution in Cygnus X$-$3. Example of a good fit solution of the theoretical model (blue solid line) to the folded {\em Fermi} lightcurve (data points). The contribution from the jet (red solid line) and the counter-jet (red dashed line) are shown for comparison. Set of parameters: $\beta=0.45,~H=3 d,~\phi_j=12^{\rm{o}},~\theta_j=106^{\rm{o}}$ and $P_e=10^{38}~\rm{erg~s}^{-1}$.}
   \label{cerutti10_fig1}
\end{figure}

\begin{figure}[h]
   \centering
   \includegraphics*[width=5.5cm]{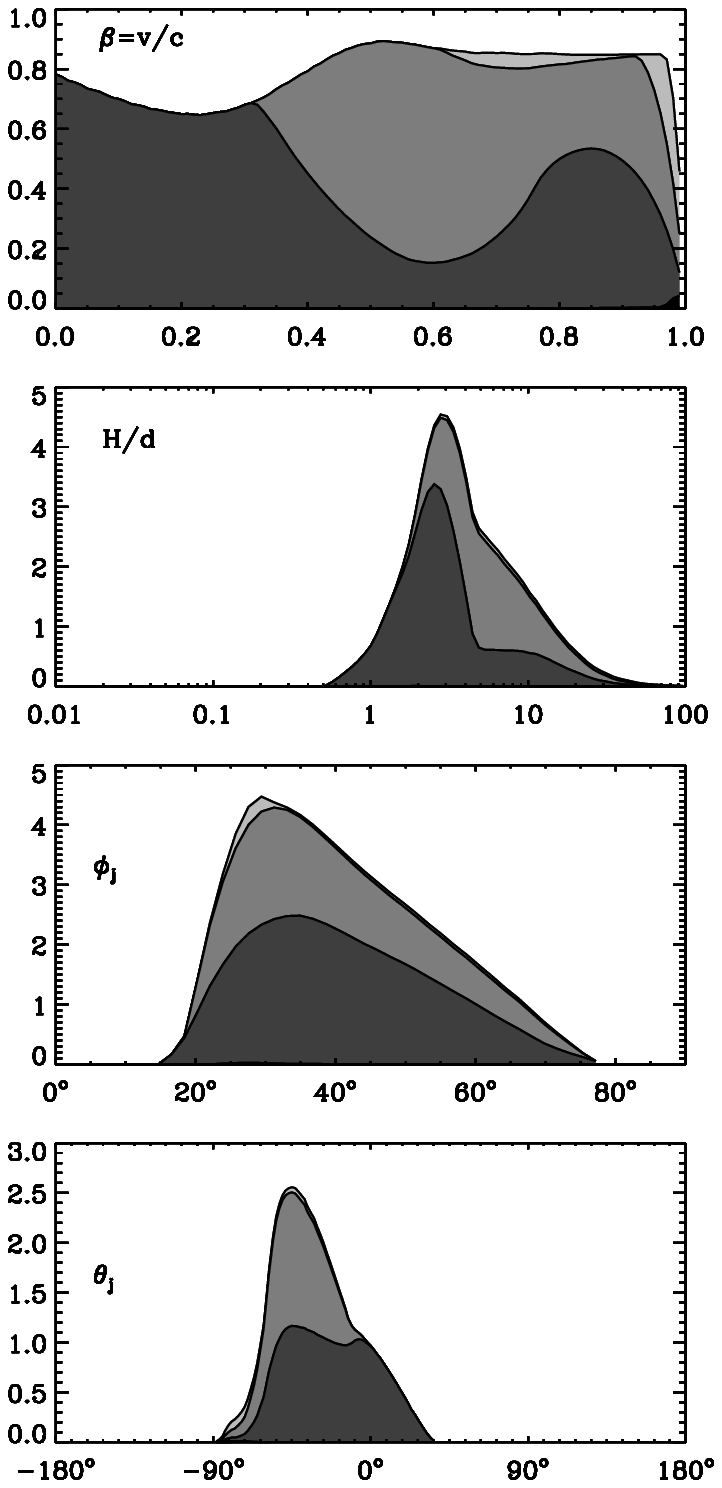}
   \includegraphics*[width=5.5cm]{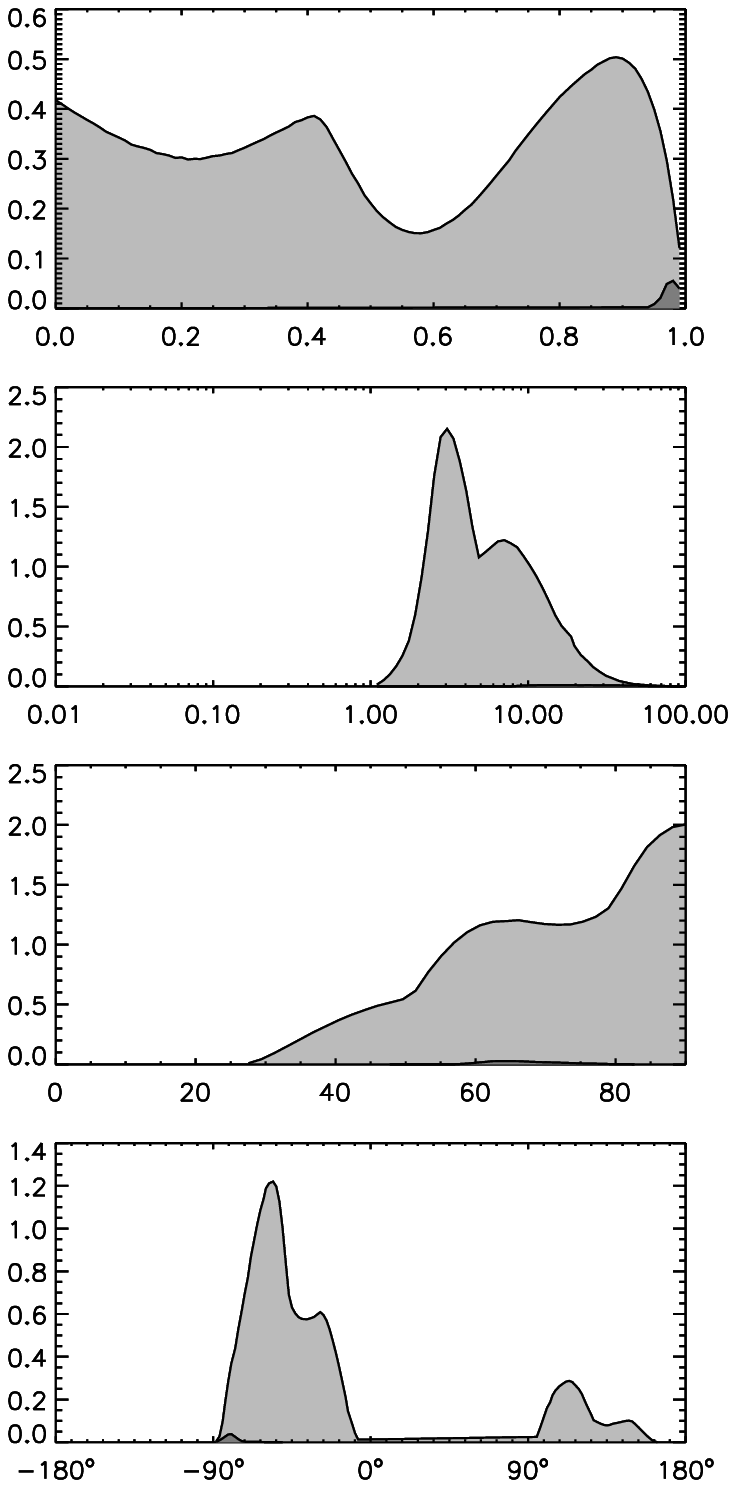}
   \caption{Distribution of good fit models contained in the 90\% confidence region of the $\chi^2$ statistics for the black hole (left panel) and the neutron star orbital solutions (right panel), for the parameters of the jet $\beta$ ({\em top} panels), $H,~\phi_j,$ and $\theta_j$ ({\em bottom} panels). Filled regions correspond to a total power injected into pairs $P_e<L_{edd}$ (light gray), $<0.1~L_{edd}$ (gray) and $<0.01~L_{edd}$ (dark gray), where $L_{edd}$ is the Eddington luminosity. 
}
   \label{cerutti10_fig2}
\end{figure}

In order to constrain the orientation of the jet, we carried out an exhaustive exploration of the space parameter. The theoretical solutions are compared with the {\em Fermi} lightcurve using a $\chi^2$ test. The best fit solutions to observations are given by those minimizing the $\chi^2$. Many sets of parameters reproduce correctly the observed gamma-ray modulation. Fig.~\ref{cerutti10_fig1} shows one possible solution. Fig.~\ref{cerutti10_fig2} presents the full distribution of models leading to a good fit, {\em i.e.} contained in the 90\% confidence region\footnote{In \citet{2010MNRAS.404L..55D}, we implicitly assumed fast cooling such that $P_e\approx\int_{\gamma_{min}}^{+\infty} K_e \gamma_e^{-p}d\gamma_e/t_{ic}$ where $t_{ic}\approx0.5(\gamma_e/10^3)^{-1}(R/d)^2~$s is the inverse Compton cooling timescale at $\gamma_{min}=10^3$ (see \S4). Unfortunately, the term $(R/d)^2$ was not properly taken into account in our calculation of the distribution of acceptable parameters. The corrected distribution allows for a greater range of solutions with electrons injected at a large distance from the compact object. The corrected Figure 3 is shown here on the left panel in Fig.~\ref{cerutti10_fig2}. The parameters of the best fit model and our conclusions are unchanged.}.

It appears from this study that the jet should be inclined close to the line of sight. The jet is mildly relativistic ($\beta< 0.9$) and pairs should not be located within the system ($0.5d<H<10d$). We favor a massive compact object in the system ({\em i.e.} a black hole) as the energy budget required to reproduce the GeV flux can be only a small fraction of the Eddington luminosity. This work reveals also that the gamma-ray modulation (amplitude and shape) is very sensitive to the polar angle $\theta_j$, {\em i.e.} if the jet precesses. The non-detection by {\em COS~B} \citep{1987A&A...175..141H} and EGRET \citep{1997ApJ...476..842M} may be the consequence of a non favorable orientation of the jet. The controversial results by {\em SAS-2} \citep{1977ApJ...212L..63L} might actually be a real detection.

\section{Gamma-ray absorption}{\label{sect_abs}}

\subsection{The model}

High-energy photons of 100~MeV-1~GeV can be absorbed by $\sim$0.1-1~keV photons. In Cygnus~X$-$3, the main source of soft X-rays could be provided by an accretion disk around the compact object. Following \citet{1997ApJ...475..534Z}, we compute the gamma-ray opacity in the thermal radiation field emitted by a standard accretion disk (optically thick, geometrically thin) in Cygnus~X$-$3. The inner radius of the disk $R_{in}$ is set at the last stable orbit. Assuming that the total luminosity of the disk is radiated in X-rays $L_{disk}\approx L_X\approx 10^{38}~\rm{erg~s}^{-1}$, the accretion rate is $\dot{M}\approx 10^{-8}~M_{\odot}~\rm{yr}^{-1}$ for the black hole solution. The source of gamma rays is assumed point-like and located above the disk at an altitude $z$. The disk is inclined at an angle $\psi$ with respect to the observer.

\subsection{Results}

\begin{figure}[h]
   \centering
   \includegraphics*[width=9cm]{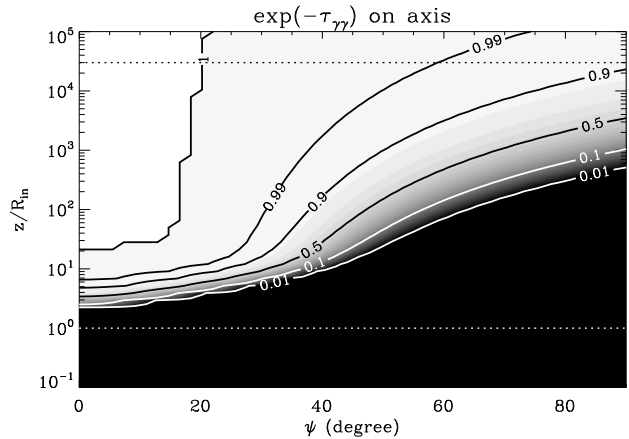}
   \caption{This map gives the gamma-ray opacity $exp\left(-\tau_{\gamma\gamma}\right)$ as a function of the viewing angle $\psi$ and the altitude of the source above a standard accretion disk $z$ for a 1~GeV gamma-ray photon. The calculation is applied here to Cygnus~X$-$3, for the black hole solution. The source is assumed to be along the axis of the disk. Black regions correspond to high opacity ($\tau_{\gamma\gamma}\gg 1$) and bright regions to low opacity ($\tau_{\gamma\gamma}\ll 1$). The inner and external radius of the accretion disk taken here are $R_{in}=10^7$~cm and $R_{ext}=10^{11}$~cm. The white dashed line indicates $z\equiv R_{in}$ and the black dotted line $z\equiv d$.}
   \label{cerutti10_fig3}
\end{figure}

Fig.~\ref{cerutti10_fig3} presents the gamma-ray opacity map of a 1~GeV photon in the radiation field of the accretion disk in Cygnus~X$-$3. Photons are injected on the revolution axis of the disk. Gamma-ray photons are highly absorbed if the source lies in a compact region around the compact object ($z<10-1000 R_{in}$). Similar maps were obtained by \citet{2010MNRAS.401.1983S} in the context of AGN with an application to Centaurus~A. In addition to the thermal component in soft X-rays, the spectrum of Cygnus~X$-$3 exhibits also a non-thermal tail in hard X-rays (see {\em e.g.} \citealt{2008MNRAS.388.1001S}). This component might be related to the emission from a hot corona of electrons above the accretion disk (see {\em e.g.} \citealt{1999ASPC..161..375C}). These photons could also contribute significantly to increase the gamma-ray opacity in the system at MeV and GeV energies. More theoretical endeavors are required in this direction.

\section{Conclusion}

Doppler-boosted Compton emission from energetic pairs accelerated at a specific location far from the compact object, in an inclined and mildly relativistic jet explains convincingly the gamma-ray modulation in Cygnus~X$-$3. The lack of absorption signature in the GeV emission implies the source is at least $10^8-10^{10}$~cm above the accretion disk. Microquasars provide a nearby and well constrained environment to study accretion-ejection mechanisms and acceleration processes in relativistic jets.

\begin{acknowledgements}
{\bf Acknowledgements:} This work was supported by the {\em European Community} via contract ERC-StG-200911.
\end{acknowledgements}

\bibliographystyle{aa}
\bibliography{cerutti10}

\end{document}